\documentclass[%
 aip,
 amsmath,amssymb,
 reprint,%
]{revtex4-1}
\usepackage{hyperref}
\usepackage{graphicx}
\usepackage{dcolumn}
\usepackage{bm}

\usepackage[utf8]{inputenc}
\usepackage[T1]{fontenc}
\usepackage{mathptmx}
\usepackage{etoolbox}
\usepackage[dvipsnames]{xcolor}
\usepackage{amsmath}

\makeatletter
\def\@email#1#2{%
 \endgroup
 \patchcmd{\titleblock@produce}
  {\frontmatter@RRAPformat}
  {\frontmatter@RRAPformat{\produce@RRAP{*#1\href{mailto:#2}{#2}}}\frontmatter@RRAPformat}
  {}{}
}%
\makeatother
\begin{document}

\preprint{AIP/123-QED}

\title[Sample title]{Geometric Criteria for Complete Mode Conversion in Detuned Systems via Piecewise-Coherent Modulation}
\author{Awanish Pandey}
 \email{opcawanish@iitd.ac.in}

 \altaffiliation{Optics and Photonics Center, Indian Institute of Technology Delhi, New Delhi 110016, India}

\date{\today}

\begin{abstract}
Static phase detuning fundamentally constrains coherent state transfer in asymmetric classical and quantum systems. We introduce a Bloch-sphere formulation for piecewise-coherent modulation that recasts coupled-mode dynamics as geometric trajectories, transforming algebraic control into path optimization. The approach reveals a cone of inaccessibility at the target pole and yields exact geodesic criteria for complete mode conversion in detuned systems. Leveraging this framework, we break time-reversal symmetry to realize a magnet-free optical isolator with near-unity contrast. Furthermore, for detuning larger than coupling between modes, we develop a recursive multi-step protocol enabling deterministic transfer for arbitrary detunings and derive a universal geometric lower bound on the required number of coupling-switching events.
\end{abstract}

\maketitle
\begin{figure*}[htbp]
\centering
 \includegraphics[width=\linewidth]{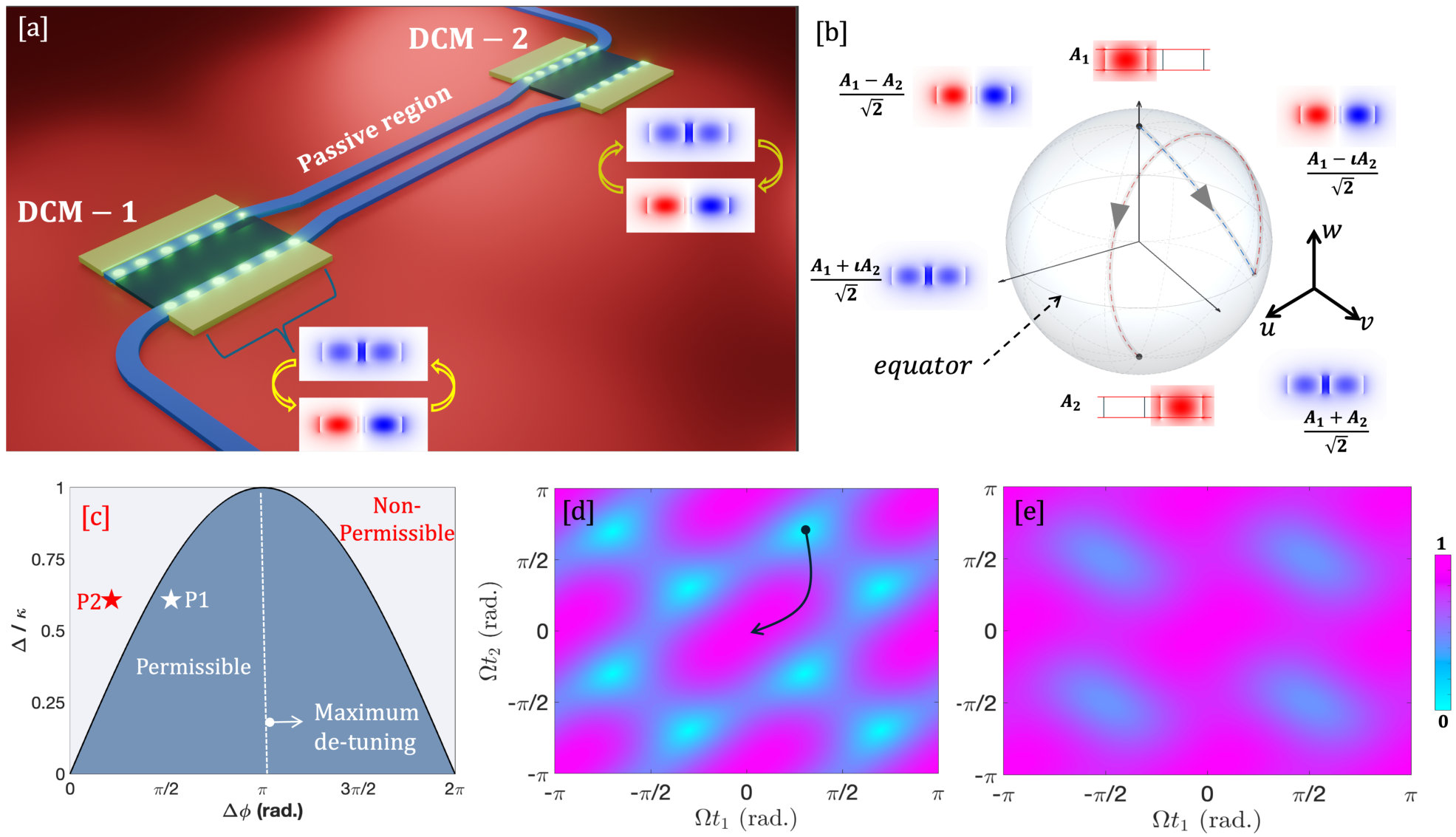}
 \caption{(a) Schematic of the directional coupler modulator (DCM) structure. First, we study the energy transfer between the supermodes in a single directional coupler modulator (DCM-1) and later, build a system with a second DCM-2 to study non-reciprocal optical transmission, [b] representation of directional coupler modes on a Bloch sphere with the curve showing the state evolution for $\Omega t_t = 0.29\pi$ and $\Omega t_2 = 0.71\pi$ leading to complete transfer of energy from $A_1$ to $A_2$, [c] permissible values of $\phi$ and $\Delta/\kappa$ for complete exchange of energy between the supermodes of the directional coupler, [d] and [e] optical energy transferred to $A_2$ supermode as a function of $\Omega_1t_1$ and $\Omega_2t_2$ for two point $P_1,P_2$ respectively as shown in [c] showcasing possible range of energy transfer between the supermodes.}\label{fig:schematic}
\end{figure*}

Tunable energy transfer between distinct eigenmodes is a cornerstone of coherent control in both quantum and classical wave physics \cite{intro1,intro2,intro3,intro4}. In static linear systems, however, this exchange is strictly governed by phase matching. Any structural asymmetry or detuning between the coupled modes creates a potential barrier that fundamentally forbids complete range of energy transfer e.g. between de-tuned qubits, acoustic resonators with resonance splitting, optical resonators with distinct resonance frequency,  or directional couplers with either nonidentical waveguides or supermodes at different frequencies \cite{detuned1,acoustic,detuned2,detuned3,detuned4}. Consequently, achieving arbitrary state control in such mismatched environments necessitates a departure from stationary interactions toward dynamic, time-varying protocols such as Floquet mode conversion  \cite{theodoros, loncar_pm, shanhui_interband,bahl_acoustic,fan_bloch,quantum1}. By tailoring the temporal profile of the drive, one can couple different modes with the help of generated sidebands that steer the system modes to any target state, effectively bypassing the constraints of the static energy landscape \cite{f1} .

Traditional continuous-wave modulation protocols rely on a strict frequency commensurability between the drive and the intrinsic mode detuning \cite{manuj,floquet1,lipson_nature}. However, such parametric drives inherently introduce a time-dependent phase accumulation that manifests as an unwanted spectral chirp in the signal field \cite{chirp1, chirp2}. Furthermore, periodic modulation inevitably couples the system to a higher-order Floquet modes, leading to energy leakage into parasitic sidebands that degrades the net transfer efficiency \cite{pandey1,popovic1}. The strict frequency-commensurability constraint fundamentally bottlenecks the protocol’s usable bandwidth, rendering it inherently incompatible with broadband signals and highly susceptible to dynamically drifting detunings \cite{f2,f3}.

In this work, we implement coherent control to enable arbitrary energy exchange in intrinsically asymmetric directional couplers. Using a piecewise-coherent modulation protocol, we identify a modulation pattern that converts the design task into geometric trajectory optimization on the Bloch sphere, explicitly compensating static phase detuning. This framework shows that sequencing interaction intervals can steer the supermodes to any desired population ratio—including unity transfer despite structural mismatch by identifying a suitable path on the Bloch sphere. In doing so, we eliminate the incomplete Rabi cycles of detuned couplers and convert partial oscillations into deterministic, full-range state transfer \cite{rabi,rabi1}. To address regimes where the detuning exceeds the coupling strength between the modes, we establish a recursive multi-step framework that enables deterministic transfer for arbitrary detunings. In this high-detuning limit, we show that the state vector "staircases" toward the target pole by optimally timing switches at intersection points where the downward velocity is maximized.

We also harness this deterministic transfer capability to explicitly break time-reversal symmetry \cite{timer1,timer2,timer3}, realizing a magnet-free optical isolator with near-unity contrast. We implement this scheme by cascading two such dynamically modulated coupling stages, separated by a passive phase shifter. We underscore that the Bloch sphere representation is not merely a tool for mapping stationary states. Rather, we establish a rigorous framework wherein the dynamic evolution of states within a directional coupler is projected onto the sphere, effectively recasting the challenge of complete energy transfer as a geometric path optimization problem. Our analysis reveals that the stability of the underlying path ensures consistent high-efficiency transfer even of nominal loss in the system, offering a significant advantage over conventional resonant approaches \cite{dutt_nonreciprocal,bhave}.

We illustrate the approach using an optical directional coupler. However, the underlying geometric analysis is universal and directly extends to detuned two-mode coupled systems across photonics, acoustics, mechanics, and quantum platforms. The schematic of the structure is shown in Fig.~\ref{fig:schematic}(a). The device consists of an asymmetric directional-coupler modulator (DCM-1) that supports an even and an odd supermode. The asymmetry may arise from unequal constituent waveguide cross-sections and/or from a modal frequency mismatch, such that the two supermodes are centered at angular frequencies $\omega_1$ and $\omega_2$. Following the coupled mode theory (CMT) treatment \cite{detuned3}, it could be shown that complete power transfer is forbidden in a detuned coupler. The maximum transferable fraction (at particular time-intervals) is bounded by $\kappa^2/(\Delta^2+\kappa^2)$, where $\kappa$ is the coupling rate and $\Delta$ denotes the detuning. In a frame rotating at $\Delta \equiv (\omega_1-\omega_2)/2$ \cite{detuned1}, the evolution for the field amplitudes associated with the modes at $\omega_1$ and $\omega_2$ are written as:

\begin{equation}
  \mathbf{A}(t)
=
\begin{bmatrix}
A_1(t) \\
A_2(t)
\end{bmatrix}
=
\begin{bmatrix}
A & -j B \\
-j B^* & A^*
\end{bmatrix}
\mathbf{A}_0
=
\mathbf{M}(t)\,\mathbf{A}_0  
\end{equation}

where 
\begin{equation}
\label{eq:A}
A(t) = \cos(\Omega t)
- j\,\frac{\Delta}{\Omega}\,\sin(\Omega t)
\end{equation}
\begin{equation}
\label{eq:B}
B(t) = \frac{\kappa}{\Omega}\,\sin(\Omega t).
\end{equation}

$\Omega = \sqrt{\Delta^2 + |\kappa|^2}$ is defined as the Rabi oscillation frequency and $A_0$ is the initial condition for the optical energy in the supermodes ranging between $[0,1]$. For arbitrary power ratios, achieving $A_2 = 1$ transfer guarantees access to any intermediate fraction between 0 and 1. Eqs.~\eqref{eq:A} and~\eqref{eq:B} show that evolution from the initial state $A_0=\begin{bmatrix}1 & 0\end{bmatrix}^{\mathsf{T}}$
to the target state $A_F=\begin{bmatrix}0 & 1\end{bmatrix}^{\mathsf{T}}$,
corresponding to complete transfer from mode $A_1$ to mode $A_2$, is possible only when the detuning vanishes, $\Delta=0$. Under this condition, full transfer occurs at times $t=\pi/(2\kappa)$.

The complex amplitudes of the two waveguide modes, $\mathbf{A}(t) = (A_1, A_2)^\mathsf{T}$, map to a real three-dimensional Bloch vector $\mathbf{S} = (u,v,w)$, defined by the parameters $u = 2\mathrm{Re}\{A_1 A_2^*\}$, $v = 2\mathrm{Im}\{A_1 A_2^*\}$, and $w = |A_1|^2 - |A_2|^2$ (see supplementary section). In this representation, the North and South poles, $\mathbf{S} = (0,0,\pm 1)$, correspond to energy localized entirely in either of the individual waveguides of the DC. Points on the equator ($w=0$) describe equal-energy superpositions, including the symmetric and antisymmetric supermodes aligned along $\pm u$ as shown in Fig. \ref{fig:schematic}b. Under static conditions with fixed detuning $\Delta$ and constant coupling $\kappa$, the system dynamics are governed by the equation of motion $\dot{\mathbf{S}} = \boldsymbol{\Omega} \times \mathbf{S}$ where $\dot{\mathbf{S}}$ is the time derivative of the Bloch vector which could be visualized as the velocity of the vector on the sphere and $\Omega$ is the rotation vector. This implies that $\mathbf{S}(t)$ undergoes a rigid precession on the Bloch sphere about a fixed torque vector defined by the unit-vector rotation axis:
\begin{equation}
\hat{\mathbf{n}} = \frac{1}{\Omega}\left(\mathrm{Re}\{\kappa\}, \mathrm{Im}\{\kappa\}, \Delta\right), \quad \Omega = \sqrt{\Delta^2 + |\kappa|^2},
\end{equation}
with a precession frequency $2\Omega$. Crucially, when $\Delta \neq 0$, the rotation axis acquires a finite $w$-component and tilts away from the equatorial plane. Consequently, a state initialized at the North pole precesses along a cone centered on $\hat{\mathbf{n}}$ that describes a latitude strictly bounded away from the South pole. Thus, a constant-$\kappa$ evolution is geometrically forbidden from achieving complete energy transfer between detuned waveguides.

Temporal modulation restores complete transfer by switching the Bloch-sphere rotation axis. For $0\le t\le t_1$, set $\kappa(t)=\kappa_1 e^{\iota\phi_1}$ switching to $\kappa(t)=\kappa_2 e^{\iota\phi_2}$ for $t_1\le t\le t_2$ reorients the axis to $\hat{\mathbf{n}}_2=\Omega^{-1}\!\left(\mathrm{Re}\{\kappa_2 ,\mathrm{Im}\{\kappa_2 \},\Delta\right)$. Depending on the device characteristics and the modulation phase $\phi (=\phi_2 - \phi_1)$, the resulting precession cone can be made to intersect the South Pole. The resulting state after the two segments is:




\begin{equation}
    A_F(t) = M_{t2}M_{t1}A_0 = \begin{bmatrix}
D & O \\
-O & D^{*}
\end{bmatrix}
A_0
\end{equation}

with $M(t_{i})$ the segment propagator for duration $t_i$ and $D,O$ being the diagonal and off-diagnoal terms of $M$ after the switching event. Imposing the $(1,1)$ element of the composite evolution to zero for the chosen input yields conditions on $t_1$, $t_2$, $\kappa{1}$, $\kappa_{2}$, $\Delta$, and $\phi$ that implement an amplitude inversion between the two modes. Since this design problem is intrinsically multi-parameter, we develop a Bloch-sphere formulation to track the mode populations as a trajectory on the sphere, thereby recasting energy transfer as a geometric control problem and assume $\kappa_1 = \kappa_2 = \kappa$. While prior frameworks have established the Bloch sphere as a robust tool for representing stationary coupled-mode states \cite{cherchi,frigo,ulrich,korotky}, we extend this mapping to the domain of coherent modulation. By implementing piecewise-coherent modulation, we recast the challenge of complete energy transfer in asymmetric systems as a trajectory optimization problem, effectively steering the state vector across topologically forbidden boundaries.

\maketitle
\begin{figure}[htbp]
\centering
 \includegraphics[width=75 mm]{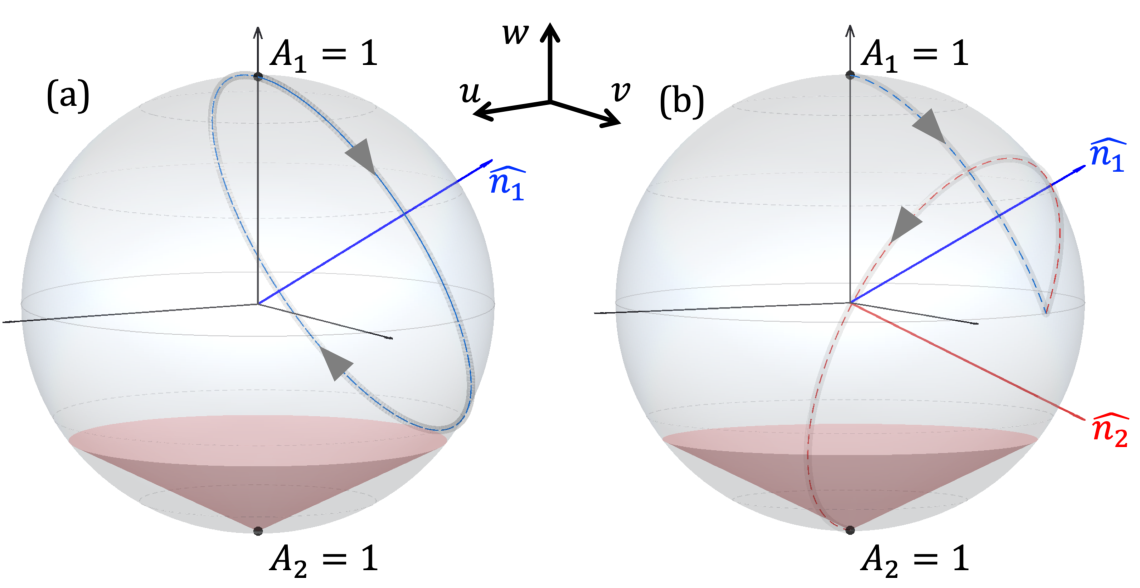}
 \caption{State evolution of supermodes on Bloch sphere in under [a] non-zero $\Delta$ with a static coupling, and [b] non-zero $\Delta$ with the precession vector changes at $t_1$ leading to a complete transfer of power to $A_2$ state at time $t_2$. The red cone depicts the cone of inaccessibility.}\label{fig:switching}
\end{figure}

The transfer protocol thus proceeds as a composite rotation: during the first interval $t_1$, the Bloch vector rotates about $\hat{\mathbf{n}}_1$ from the North pole to an intermediate point on the sphere surface. In the subsequent interval $t_2$, the axis switches to $\hat{\mathbf{n}}_2$, allowing the trajectory to continue along a new arc.  To force the $(1,1)$ component of the final evolution matrix to vanish (thereby achieving the target state $S=(0,0,-1)$), the distinct circular orbits defined by the rotation axes $\mathbf{\hat{n}}_1$ and $\mathbf{\hat{n}}_2$ must physically intersect on the sphere. This requirement can be derived from spherical geometry. The first rotation (starting from the North pole) traces a circle centered on $\mathbf{\hat{n}}_1$ with an angular radius $\rho_1 = \frac{\pi}{2} - \psi$, where $\psi = \arctan(\Delta/|\kappa|)$ is the axis elevation. Conversely, for the second rotation to converge to the South pole, the switching point must lie on a trajectory centered on $\mathbf{\hat{n}}_2$ with an angular radius $\rho_2 = \frac{\pi}{2} + \psi$.

For these two circular loci to share a common switching point, the geodesic angle $\Theta_\Omega$ between the rotation axes must satisfy the spherical triangle inequality $\Theta_\Omega \geq |\rho_2 - \rho_1|$ ensuring that the two circles intersect. Substituting the angular radii yields the fundamental geometric constraint $\Theta_\Omega \geq \left( \frac{\pi}{2} + \psi \right) - \left( \frac{\pi}{2} - \psi \right) = 2\psi.$ This inequality reveals the physical origin of the transfer limit: $2\psi$ defines the angular aperture of a cone of inaccessibility centered at the South pole. Under static detuning, trajectories are confined to latitudes above this cone and hence, transfer is achievable if and only if the switching event steers the axis by an angle $\Theta_\Omega$ sufficient to bridge this gap. To evaluate this condition analytically, we calculate the dot product of the normalized axes $\mathbf{\hat{n}}_1$ and $\mathbf{\hat{n}}_2$, which share a common magnitude $\Omega = \sqrt{\kappa_0^2 + \Delta^2}$ and differ only by the azimuthal phase shift $\phi$ given by:
\begin{equation}
\label{eq:inequality1}
    \cos(\Theta_{\Omega}) = \mathbf{\hat{n}}_1 \cdot \mathbf{\hat{n}}_2 = \frac{\kappa_0^2 \cos(\phi) + \Delta^2}{\kappa_0^2 + \Delta^2}.
\end{equation}
Using the identity $\tan\psi = \Delta/\kappa_0$, we express the double angle as $\cos(2\psi) = (\kappa_0^2 - \Delta^2)/(\kappa_0^2 + \Delta^2)$. Since the cosine function is monotonically decreasing on $[0, \pi]$, the inequality $\Theta_\Omega \geq 2\psi$ implies:

\begin{equation}
    \label{eq:inequality}
    \cos(\Theta_\Omega) \leq \cos(2\psi)
\end{equation}


The inequality defines the exact angular gap $(2\psi)$ that the switching event must bridge to link the two distinct trajectories. When it is satisfied, the corresponding circular trajectories necessarily intersect, enabling the state vector to transition onto the trajectory that reaches the target south pole as shown with the state progression in Fig. \ref{fig:schematic}b. This geometric requirement places a constraint on the phase of the modulation. In particular, eqn. \ref{eq:inequality1} and \ref{eq:inequality} shows that achieving the largest detuning tolerance occurs for $\phi=\pi$, which requires the coupling to change sign between the two time intervals (push-pull modulation), as illustrated in Fig.~\ref{fig:schematic}c. Since in both intervals $|\kappa_{1,2}|=\kappa$, only the sign flips making it a push-pull type modulation. The variation of optical energy in $A_2$ as a function of $t_1$ and $t_2$ is shown in Fig.~\ref{fig:schematic}d and e for two values of $\phi$ and $\Delta/\kappa$ depicted as point $P_1$ ($\frac{\Delta}{\kappa} = 0.5, \phi = \frac{\pi}{4}$) and $P_2$ ($\frac{\Delta}{\kappa} = 0.5, \phi = \frac{\pi}{2}$), respectively, in Fig.~\ref{fig:schematic}c. In Fig.~\ref{fig:schematic}c, the black arrow represents the evolution of the state trajectory on the Bloch sphere shown in Fig.~\ref{fig:schematic}b. For $\Omega t_1 = 0.29\pi$ and $\Omega t_2 = 0.71\pi$, this evolution results in complete transfer of optical power from $A_1$ to $A_2$ implying a full trajectory from the North Pole to the South Pole is traversed on the Bloch sphere. As evident, outside the permissible range of these values ($P_2$), complete transfer of optical energy to $A_2$ is forbidden (Fig. \ref{fig:schematic}e).


Building on the arbitrary-range transfer protocol, we extend the same coherent control framework to engineer non-reciprocal optical transmission. The non-reciprocal element comprises two dynamically modulated directional-coupler stages ($M_1$, $M_3$) cascaded around a passive phase-delay region ($M_2$), as depicted in Fig.~\ref{fig:schematic}(a). In the forward/backward paths, the modulation in $M_3$ is applied with an additional phase shift $\delta$ relative to $M_1$, and the passive section accumulates a differential supermode phase $\Delta\theta=\theta_1-\theta_2$. The system dynamics are governed by the single-stage transfer matrix ${M}$, derived from the piecewise-coherent evolution with diagonal ($A$) and off-diagonal ($B$) elements satisfying $|A|^2 + |B|^2 = 1$. The total transfer matrices for forward (${T}_{\rightarrow} = {M}_3 {M}_2 {M}_1$) and backward (${T}_{\leftarrow} = {M}_1 {M}_2 {M}_3$) propagation are constructed by sequencing the individual operations. Specifically, the additional phase shift in the third stage scales the off-diagonal coupling terms as $O \to O e^{-i\delta}$, while the passive region is diagonal, ${M}_2 = \mathrm{diag}(e^{i\phi_1}, e^{i\phi_2})$.
The breaking of reciprocity manifests in the cross-mode transmission coefficients, $T_{12}$ (input mode 2 to output mode 1). The coherent interference between the constituent stages yields distinct complex amplitudes for counter-propagating signals:
\begin{align}
({T}_{\rightarrow})_{12} &= D O e^{i\phi_a} + O D^* e^{i(\phi_b - \delta)}, \\
({T}_{\leftarrow})_{12} &= D O e^{i(\phi_a - \delta)} + O D^* e^{i\phi_b}.
\end{align}
While the individual magnitudes of the matrix elements are identical, the relative phase difference between the two interference paths depends on the propagation direction. The resulting power transmission contrast is given by:
\begin{equation}
|({T}_{\rightleftarrows})_{12}|^2 = 2 |D|^2 |O|^2 \left[ 1 + \cos(\Delta\theta \pm \delta) \right],
\label{eq:transmission}
\end{equation}
where the positive sign corresponds to the forward direction. Equation~(\ref{eq:transmission}) reveals the fundamental mechanism of isolation: the transport efficiency is modulated by the sum ($\Delta\theta + \delta$) or difference ($\Delta\theta - \delta$) of the spatial and temporal phases. Consequently, whenever the differential phases are non-trivial ($\Delta\theta, \delta \neq n\pi$), the system becomes non-reciprocal. Maximum isolation is achieved by tuning the phases such that $\Delta\theta + \delta = \pi$ (destructive interference, blocking forward transport) while $\Delta\theta - \delta = 0$ (permitting backward transport), or vice versa.

In the forward propagation direction, the non-reciprocal transmission relies on the precise geometric coordination between the intermediate static phase shift ($\Delta\theta$) and the dynamic RF phase offset ($\delta$). The first modulation stage (DCM-1) rotates the state vector from the North Pole to an intermediate latitude, leaving it at a specific longitude determined by the interaction time. The static phase shift $\Delta\phi$ then plays the critical role of a longitudinal bridge. It rigidly rotates the state vector azimuthally around the polar axis ($w$-axis) without altering its energy distribution. This rotation is necessary to align the state with the intake manifold of the second modulation stage (DCM-2). The additional phase offset $\delta$ determines the azimuthal orientation of the rotation axis $\hat{n}_3$ for this second stage as shown in Fig. \ref{fig:non_rec}a. For forward transmission, the system is tuned such that the sum of the spatial and temporal phases satisfies the alignment condition (e.g., $\Delta\theta + \delta = 0$ or $2\pi$). This ensures that the azimuthal rotation $\Phi$ carries the state vector exactly to the starting longitude of the arc defined by $\delta$, allowing the two distinct rotations to stitch together into a single, continuous trajectory that descends all the way to the South Pole (complete energy transfer).

\maketitle
\begin{figure}[htbp]
\centering
 \includegraphics[width=75 mm]{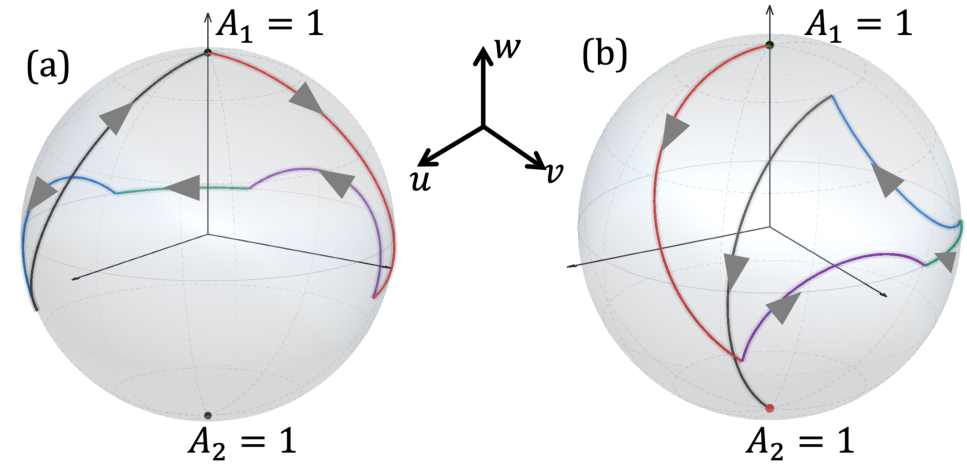}
 \caption{State evolution of supermodes on Bloch sphere in [a] forward direction, and [b] reverse direction.}\label{fig:non_rec}
\end{figure}


Conversely, in the backward direction, the interplay between $\Delta\theta$ and $\delta$ fundamentally breaks this geometric alignment due to the failure of time-reversal symmetry. While the passive region imparts the same longitudinal rotation $\Phi$ regardless of direction, the effective sign of the temporal phase offset $\delta$ relative to the spatial phase is inverted for the backward wave. This results in a phase relationship (e.g., $\Delta\theta - \delta$) that fails the alignment condition. Geometrically, the bridge $\Delta\theta$ now rotates the state vector to a longitude that is completely mismatched with the rotation axis of the first stage (which acts as the second stage for the backward wave) as shown in Fig. \ref{fig:non_rec}b. Instead of continuing the descent, the mismatched rotation axis forces the state vector onto a disjointed trajectory that effectively reverses the initial excitation. This manifests as destructive interference where the cross-coupling terms cancel out, confining the trajectory to the northern hemisphere and resulting in near-unity isolation. The device thus functions by converting a direction-dependent geometric phase mismatch directly into a topological barrier on the Bloch sphere.


It should be noted that if either the static phase shift $\Phi$ or the modulation phase offset $\delta$ vanishes, non-reciprocal transmission is fundamentally forbidden. On the Bloch sphere, setting either parameter to zero preserves latitudinal symmetry between the forward and backward trajectories. The isolation mechanism relies critically on the interference term $\cos(\Delta\theta \pm \delta)$. When either parameter vanishes, the directional distinction collapses. From a geometric perspective, the absence of either spatial ($\Delta\theta$) or temporal ($\delta$) symmetry breaking implies that the backward trajectory is simply a longitudinal mirror of the forward path. Both trajectories therefore traverse equivalent arcs on the Bloch sphere and terminate at the same final latitude (energy), thereby fully restoring reciprocity.

Finally, we mitigate the fundamental limitation of the two-step piecewise-coherent modulation protocol, where complete energy transfer is strictly bounded by the condition that the detuning $\Delta$ cannot exceed the coupling strength $\kappa$. We achieve it by introducing a $N$ step piecewise modulation protocol. In the geometric representation on the Bloch sphere, achieving deterministic state transfer requires that the sequence of precession circles, defined by the rotation axes $\hat{n}_i$, satisfies a recursive intersection requirement. This constraint dictates that for any $N$-step protocol, the circular loci of the $i$-th and $(i+1)$-th interaction stages must physically overlap to allow the Bloch vector to transition between manifolds. As discussed, for a detuned system, the rotation axis is tilted from the equatorial plane by an elevation angle $\psi = \arctan(\Delta/|\kappa|)$, creating a cone of inaccessibility at the South Pole with an angular aperture of $2\psi$. To bridge this gap when $\Delta > \kappa$, we establish the recursive intersection formula: $\Theta_{\Omega,i} \ge |\rho_{i+1} - \rho_i|$, where $\Theta_{\Omega,i}$ is the geodesic angle between successive rotation axes and $\rho$ is the angular radius of the precession. By sequencing interaction intervals such that this inequality holds at every switching event, we can ensure that the final state reaches to the south pole.

\maketitle
\begin{figure}[htbp]
\centering
 \includegraphics[width=0.8\linewidth]{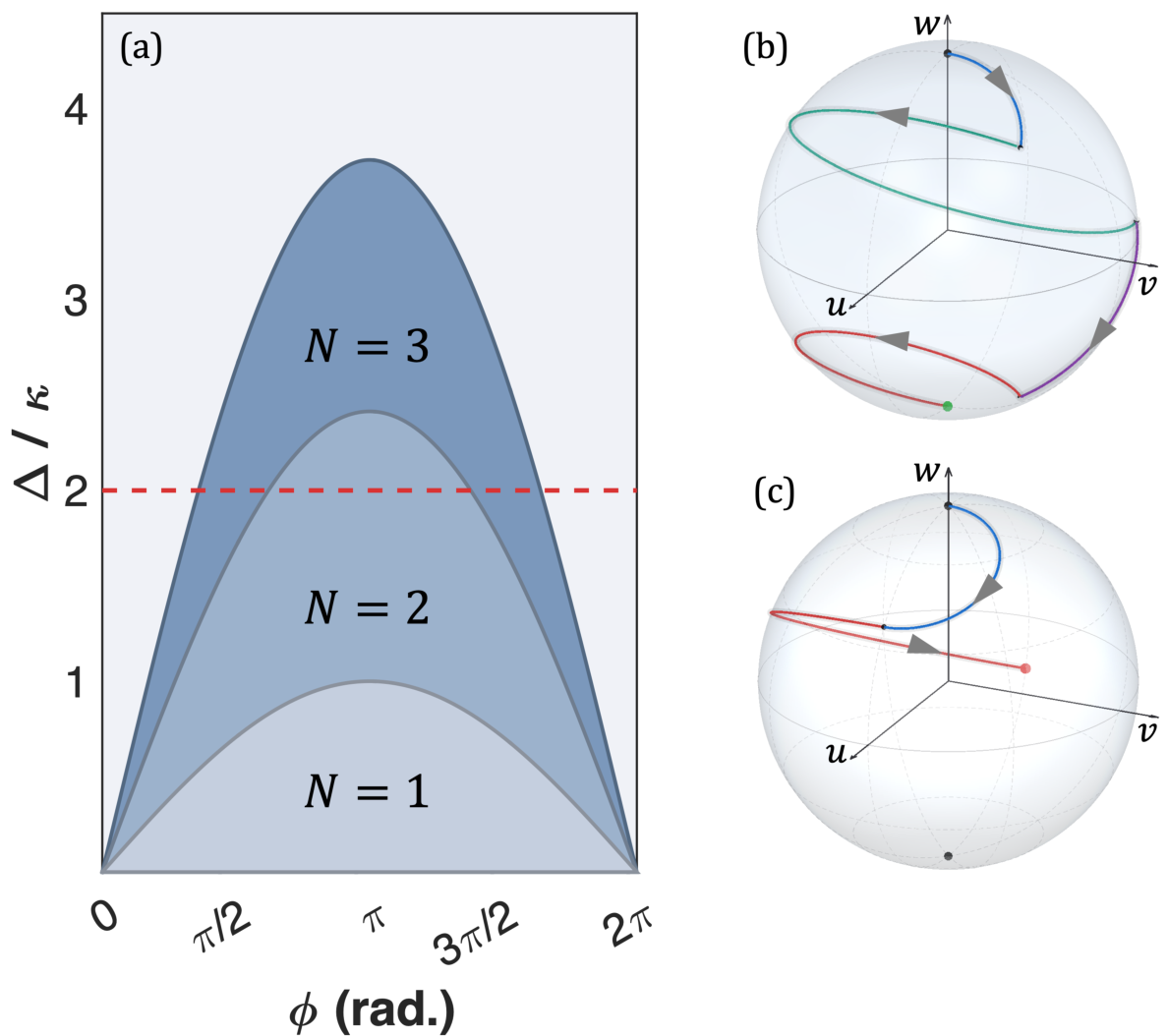}
 \caption{(a) Variation of number of coherent modulation steps required for complete state tranefer from the north to south pole, (b) complete conversion for $\Delta/\kappa=2$ using $N=2$, and (c) inability of the two step piecewise protocal for complete transfer.}\label{fig:multiple_switching}
\end{figure}

To deduce the minimum number of switching events $M_{min}$ from this recursive framework, we analyze the cumulative vertical progress made as the state vector steps down the sphere from the north pole to the south pole. In the limit of high detuning where $\Delta \gg \kappa$, the rotation axis $\hat{n}$ resides very close to the North Pole, and the angular radius of each precession circle shrinks to $\rho \approx \kappa/\Delta$. The maximum vertical displacement toward the South Pole attainable in a single interaction interval is restricted by the diameter of this small orbit, $2\rho$. However, by optimally timing the switches to occur at the intersection points where the downward velocity is maximized, we could stitch these arcs into a monotonic descent. The total geodesic distance of $\pi$ must be covered by the sum of these incremental vertical shifts, leading to the geometric requirement $\sum_{i=1}^{M} 2\rho_i \approx \pi$. Utilizing the small-angle approximation of the algebraic bound $\Delta/\kappa_0 = \tan(\pi/4N)$, we derive the fundamental limit $N_{min} \approx \pi\Delta / 4\kappa$ as shown in Fig. \ref{fig:multiple_switching}a. This result signifies that each switching event provides an effective vertical "step size" of $4\kappa/\Delta$ on the sphere. Consequently, while a single switch fails to bridge the $2\psi$ gap when $\Delta > \kappa$, a cascaded sequence of $M$ steps satisfying the recursive intersection condition enables deterministic, near-unity transfer for arbitrary detunings as shown in Fig. \ref{fig:multiple_switching}b for $\Delta/\kappa = 2$ while under a single switching, it has not reached the south pole as shown in Fig. \ref{fig:multiple_switching}c.

In conclusion, we have established a universal geometric framework for achieving arbitrary state transfer in intrinsically asymmetric coupled-mode systems. By elucidating the topological barriers imposed by static detuning—visualized as a cone of inaccessibility on the Bloch sphere—we demonstrated that piecewise-coherent modulation can steer supermodes along deterministic trajectories that bypass these limits. This control strategy proves to be more than a remedy for impedance mismatch; it is a mechanism for explicit symmetry breaking. By cascading dynamic coupling stages with passive phase delays, we introduced a magnet-free optical isolator where non-reciprocity arises purely from the direction-dependent interference between spatial and temporal geometric phases. The resulting near-unity contrast and resilience to modulation imperfections confirm that the topological stability of the geometric path translates directly to robust device performance. Our findings thus bridge the gap between temporal coupled-mode theory and geometric phase control, providing a scalable, material-independent paradigm for non-reciprocal signal processing in next-generation integrated photonics

\medskip
\textbf{Disclosures} The author declares no conflict of interest.

\textbf{DATA AVAILABILITY}
The data that support the findings of this study are available
from the corresponding author upon reasonable request.

\bibliography{main_biblio}

\clearpage

\end{document}